\documentclass[journal]{IEEEtran}

\usepackage{multirow}
\usepackage{color}
\usepackage{setspace}
\usepackage{clrscode}
\usepackage{amsthm}
\usepackage{pict2e}
\usepackage{amsmath}
\usepackage{amssymb}
\usepackage{graphicx}
\usepackage{epsfig}
\usepackage{bbold}
\usepackage{algorithm}
\usepackage{algorithmic}
\usepackage{caption}
\usepackage{subcaption}
\usepackage{mathabx}
\usepackage[ps2pdf,
bookmarks=false,
bookmarksnumbered=false, 
bookmarksopen=false, 
colorlinks=false]{}

\DeclareMathOperator*{\argmax}{argmax}

\newcommand{\pvec}{\textbf{p}}

\newcommand{\mS}{\mathcal{S}}
\newcommand{\mR}{\mathcal{R}}
\newcommand{\mA}{\mathcal{A}}

\newcommand{\E}{\mathbb{E}}

\allowdisplaybreaks[2]

\begin{document}
\title{Robust and Decentralized Reinforcement Learning for UAV Path Planning in IoT Networks}

\author{\IEEEauthorblockN{Xueyuan Wang and M. Cenk Gursoy}
\thanks{Xueyuan Wang is with School of Computer Science and Artificial Intelligence, Changzhou University, Changzhou 213164, China (email: xywang@cczu.edu.cn).
	
	M. Cenk Gursoy is with the Department of Electrical
	Engineering and Computer Science, Syracuse University, Syracuse, NY, 13244
	(e-mail: mcgursoy@syr.edu).}}
\maketitle

\begin{abstract}
	Unmanned aerial vehicle (UAV)-based networks and Internet of Things (IoT) are being considered as integral components of current and next-generation wireless networks.
In particular, UAVs can provide IoT devices with seamless connectivity and high coverage and this can be accomplished with effective UAV path planning.
	In this article, we study robust and decentralized UAV path planning for data collection in IoT networks in the presence of other noncooperative UAVs and adversarial jamming attacks. We address three different practical scenarios, including single UAV path planning, UAV swarm path planning, and single UAV path planning in the presence of an intelligent mobile UAV jammer.  We advocate a reinforcement learning  framework for UAV path planning in these three scenarios under practical constraints. The simulation results demonstrate that with learning-based path planning, the UAVs can complete their missions with high success rates and data collection rates. In addition, the UAVs can adapt and execute different trajectories as a defensive measure against the intelligent jammer.
\end{abstract}

\section{Introduction}

With the relative maturity and commercialization of fifth generation (5G) cellular networks, the focus of both academia and industry is shifting towards the design and development of sixth generation (6G) wireless communication networks \cite{yang20196g}.  Among different aspects of 6G systems, unmanned aerial vehicle (UAV)-based aerial networks and Internet of Things (IoT) are attracting increasing attention since they can enable remote access, control, and monitoring \cite{mao2021optimizing}. In particular, as an emerging technology, IoT is expected to offer promising solutions to transform the
operation and role of many existing industrial systems, e.g., in transportation and manufacturing.
It has been reported that the number of IoT connections will grow 2.4-fold to 14.7 billion by 2023 \cite{mao2021optimizing}.
Hence, IoT networks typically involve large number of devices and generate, in aggregate, significant amount of data, ranging from multimedia data such as video, images, and sounds, to structured data, such as temperature, vibration, and luminous flux information \cite{li2018learning}. Most communication scenarios in IoT networks require high coverage and massive connectivity to support such large number of IoT devices, and this presents challenges in current terrestrial networks due to fading and path loss in communication links, as well as stringent resource constraints at IoT devices. UAVs have recently emerged as key enablers of seamless wireless connectivity, and can be deployed to form non-terrestrial networks to address these challenges. More specifically, the flexibility in effective trajectory planning allows UAVs to adapt their movements based the demand arising from IoT devices, improving the overall network performance.  Thus, UAV-based aerial communication networks have become critical for fast response and high coverage especially in harsh and difficult-to-reach IoT environments \cite{you2021towards}.

In the 6G vision, as yet another advance, AI and machine learning are being explored to orchestrate and manage intelligent networks \cite{araniti2021toward, saad2019vision}. In particular, machine leaning is expected to immensely benefit 6G networks, boost the performance levels while making efficient use of scarce resources, and manage complex interactions.
For instance, in UAV communications, the high mobility of UAVs may lead to frequent handovers, uncertainties in their precise locations, complications in airspace management and collision avoidance, and challenges in dynamic resource allocation. In this setting, deep reinforcement learning (DRL) can address the complex decision-making tasks in UAV networks \cite{yang2020artificial}, and
provide enhanced performance in dynamic path planning scenarios. For example,  Liu et al. considered the joint trajectory design and power control for multiple UAVs to maximize the instantaneous transmission rate of mobile users.  An echo-state network-based algorithm  was used to predict the user's movement, and  multi-agent Q-learning algorithm was used  to determine UAVs' initial positions, trajectories and power allocation \cite{liu2019trajectory}. Based on map centering, global and local map processing, Bayerlein et al. developed a double deep Q network (DDQN) to solve the UAV path planning problem under time and collision avoidance constraints to maximize the collected data from IoT nodes \cite{bayerlein2021multi}. Considering a scenario where vehicles and UAVs collaborate to collect data from multiple IoT nodes, Li et al. utilized a genetic algorithm to plan paths for vehicles, and used asynchronous advantage actor-critic (A3C)  to plan paths for energy-constrained UAVs \cite{li2021drlr}. Chen et al. proposed a constrained  multi-agent Q-learning algorithm to solve the time-constrained target search problem through autonomous flight control of UAV swarms \cite{chen2020autonomous}. Hu et al. proposed a value decomposition-based reinforcement learning algorithm combined with a meta-learning mechanism to maximize the proportion of served users by optimizing the trajectories of multiple UAV base stations \cite{hu2021distributed}. Zhong et al. proposed a mutual deep Q-network to determine the optimal three dimensional trajectory and power allocation strategy for UAVs, and used the experience of multiple agents to train a shared neural network  to shorten the training time \cite{zhong2021multi}. The above-mentioned studies have addressed the multi-UAV path planning problem in specific scenarios with good performance. However, they have not considered the fact that there are typically unknown number of noncooperative UAVs in the environment, and how to address path planning in their presence with effective collision avoidance is critical. Furthermore, there may even be adversarial attackers, e.g., performing jamming attacks on the air-to-ground links. Robust decision making under such attacks is also a crucial undertaking for both UAV operation and IoT networking.

In this article, motivated by these considerations, we consider UAV path planning for data collection in IoT networks in the presence of multiple unknown and noncooperative UAVs, which may have different missions and policies and do not communicate.
The main contributions can be summarized as follows. First, we construct three different realistic scenarios, including single UAV path planning, UAV swarm path planning, and single UAV path planning in the presence of an intelligent mobile jammer (which is itself a UAV). Then, we propose decentralized  deep reinforcement learning algorithms for UAV path planning in these three scenarios under multiple constraints, including mission completion deadline, collision avoidance, kinematic, and  connectivity constraints. The proposed algorithms operate in a decentralized fashion, and thus are able to address scenarios with different number of UAVs  as well as with noncooperative behavior. We note that the algorithms proposed for the last two scenarios are derived from the algorithm designed for single UAV path planning. Thus, we also provide insights on how to directly modify the algorithms developed for relatively simple settings to operate in more complex scenarios.

\section{UAV-Assisted IoT Networks}
\begin{figure}
	\centering
	\includegraphics[width=0.5\textwidth]{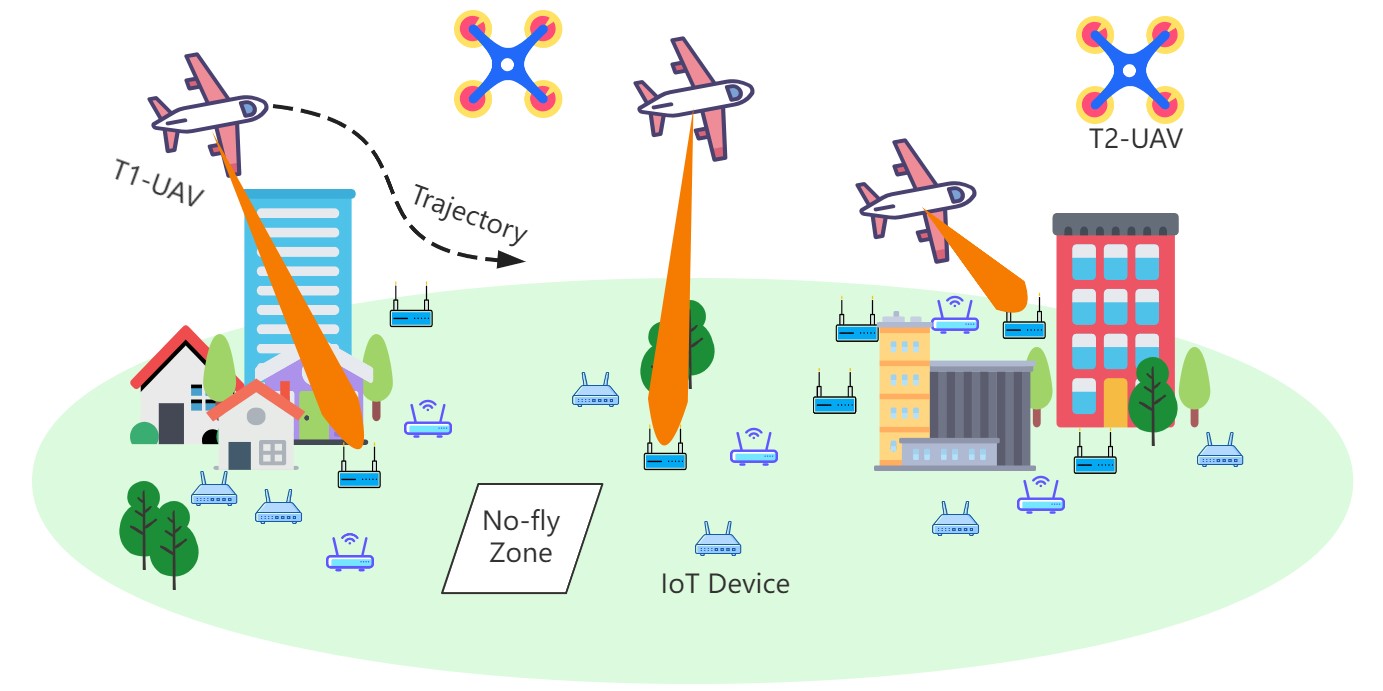}
	\caption{\small An illustration of data collection in a noncooperative multi-UAV scenario.   \normalsize}
	\label{Fig:network}
\end{figure}

An illustration of data collection from ground IoT devices in a noncooperative multi-UAV scenario is provided in Fig. \ref{Fig:network}. Note that we consider a two-layer structure with a IoT device layer and a UAV layer. The detailed descriptions of these layers are provided below.

\subsection{IoT Layer}
In the IoT layer, we have a large number of IoT devices, supporting various applications such as smart city, home intelligence, health care, remote monitoring, autonomous systems, etc. The memory and computational capacity of each IoT device is limited, and thus data typically needs to be uploaded periodically via uplink transmission. Generally, IoT devices are located at fixed ground positions and transmit at low power levels due to limited battery capacity.
We assume that the IoT devices can be in two modes: active mode, if the device still has data to transmit; and silent mode, if data upload is completed.

\subsection{UAV Layer}
In the considered multi-UAV scenario, multiple UAVs are deployed to collect data from multiple distributed ground IoT devices. When given a mission, these UAVs depart from a specific area, fly over the IoT devices, and finally arrive at the landing area.
Due to the transmit power limitation at the IoT devices and the path loss and fading, UAVs need to fly relatively close to the IoT devices to establish reliable communication links.

In the environment, there may exist noncooperative UAVs with different  missions, destinations, movements, and decision-making policies.  These UAVs do not communicate, and thus information related to their missions, trajectories, and decision making is generally unavailable. In order to establish a clear distinction, UAVs with missions to collect data from IoT nodes are denoted as T1-UAVs, and the above-mentioned noncooperative UAVs are denoted as T2-UAVs.  In the considered multi-UAV scenario, T1-UAVs work collaboratively, while T2-UAVs, as also noted above, are noncooperative, 

UAVs on a data-collection mission operate subject multiple constraints detailed below:
\begin{itemize}
	\item \emph{Power limitations:} UAVs consume energy for flying/hovering and communication, and are generally powered by onboard batteries. Battery capacity affects UAVs' performance and flight duration. 
	\item \emph{Collision avoidance:} In scenarios involving multiple UAVs, a fundamental challenge is to safely control the interactions with other dynamic agents in the environment and avoid collisions.  In addition,   UAVs should be able to navigate while staying free of collisions with fixed obstacles (such as buildings) and avoiding no-fly zones.
	\item \emph{Mission completion deadlines:} In practical applications, a UAV with a mission has to complete the required tasks within a specific time duration.
	\item \emph{Kinematic constraints:} In practice, UAVs have speed, acceleration, and angular velocity constraints during flight. In addition, different limitations should be taken into account for fixed-wing and rotary-wing UAVs.
	\item \emph{Connectivity constraints:} Different access techniques, e.g., time division multiple access (TDMA) or frequency division multiple access (FDMA), can be utilized for communication between the UAVs and the IoT devices. These lead to different connectivity constraints. For example, if TDMA is employed, UAV can communicate with at most one device at each time.
	\item \emph{Service requirements:} Depending on the application, IoT devices may have different requirements, e.g., in terms of quality of service (QoS), latency, and priority.
\end{itemize}

\section{Single UAV Path Planning} \label{Sec:single}
In this section, we consider a scenario in which a single T1-UAV is assigned a mission to collect data from IoT devices, and multiple T2-UAVs with unkown  missions exist in the environment. The goal of the T1-UAV is to optimize its path in order to maximize the collected data from all devices subject to multiple constraints, including power, collision avoidance, mission completion deadline, kinematic, TDMA, and  starting-point and end-point constraints. In addition, power constraint restricts the UAV's flight duration, and thus we convert the power limitation into time limitation which can be combined with mission completion deadline constraint. The optimization problem for T1-UAV path planning can be expressed as
 \begin{align}
(\text{P} 1):\argmax_{ \{\pvec_t,  \forall t\}}  & \qquad \sum_{t=0}^{T}\sum_{n=1}^{N} q_{nt}\Delta t R_{nt} \notag  \\
\text{subject to}          &\qquad \text{time, collision avoidance, kinematic,} \notag
\\
&\qquad \text{and connectivity constraints},	\notag
\vspace{-.4cm}
\end{align}
where the subscript $t$ is used to indicate the time step, $T$ is the total time duration, $N$ is the number of IoT devices, $q_{nt}$ indicates the association with IoT devices at time step $t$, and $R_{nt}$ is  the information rate at time step $t$. $\pvec_t$ is the location of the T1-UAV at time step $t$, and thus $\{\pvec_t,  \forall t\}$ represents the UAV trajectory.

Considering the optimization in (P1), we can translate the considered sequential decision making problem into a Markov decision process (MDP) represented by the tuple $\langle \mS,\mA,\mR \rangle$, which is described in detail below.
\subsection{State ($\mS$)}
Each UAV's information  consists of its position, velocity, radius of its body,  destination, maximum speed, and orientation, among which the position, velocity, and radius are observable and form an observable information vector, and destination, maximum speed and orientation are hidden (from other UAVs) and can form a hidden vector.
We note that unmanned vehicles/systems are typically equipped with sensors (including e.g., laser ranging, radar, electro-optical, infrared, or thermal imaging sensors, and motion detectors) for safe and robust operation. UAVs, equipped with one or two types of such sensors, will have the capability to detect the observable state of the nearby UAVs at low cost. Therefore, it can be  assumed that the T1-UAV is equipped with sensors, with which it is able to sense the existence of T2-UAVs that are closer than a  certain distance.
In addition, T1-UAV is assigned to collect data from a number of IoT devices, and thus information of these devices (e.g., their locations) can be obtained.

Therefore, the UAV's state consists of i) T1-UAV's own full information vector; ii) the observable information vector of T2-UAVs within T1-UAV's sensing region (indicating that the total number of observed T2-UAVs varies over time); iii) the location information of each IoT device, the association, the received signal power, and  the amount of remaining data at each device, which can be obtained from initial amount of data and the transmit power of the device; and iv) the available time left for the given mission.
To aid the UAV in interpreting the large state space  and obtaining more information from the state, each state vector should be parameterized before training.

\subsection{Action ($\mA$)}
Kinematic constraints restrict the UAV's speed, acceleration and rotation within a given time duration. Thus, the action space should be built based on the kinematic constraints. In this article, we assume that permissible velocities satisfying the speed and rotation constraints can be sampled to build a velocity-set. The indices of the velocities in the velocity-set constitute the action space.

\subsection{Reward ($\mR$)} \label{subsec:reward}
Reward function should be designed to encourage the T1-UAV to achieve its objective while satisfying the constraints. Therefore, the reward function consists of multiple terms, including: i) a term corresponding to the amount of data to be collected in the next time duration in order to encourage the UAV to collect more data from IoT devices; ii) a term introduced to penalize collision with other UAVs and encourage the T1-UAV to stay sufficiently away from  other UAVs; iii) a term to penalize the collision with the fixed obstacles or penalize entering into no-fly zones; iv) a term  related to the mission completion deadline constraint in order to encourage the UAVs to arrive at their destinations within the allowed time duration; v) a reward given for arriving at its destination; and vi) a step penalty to encourage fast arrival.
Note that the priority of each term can be adjusted with weights to adapt to different mission requirements.

\begin{figure}
	\centering
	\includegraphics[width=0.4\textwidth]{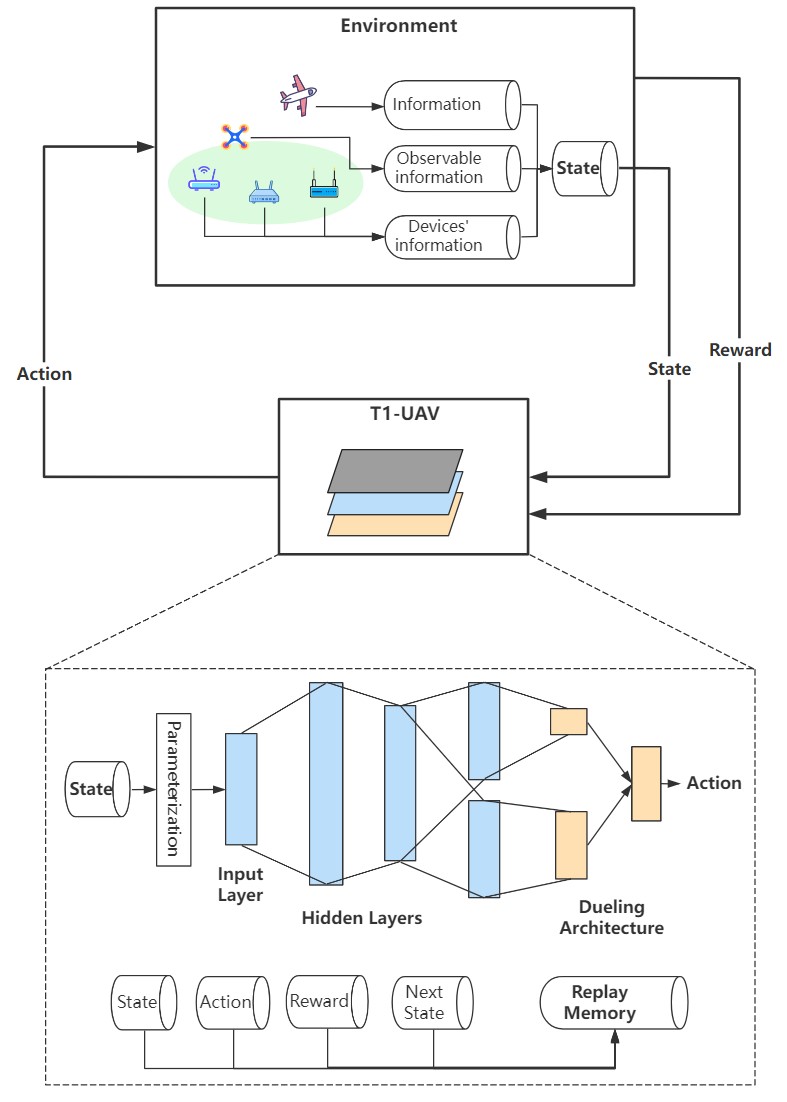}
	\caption{\small An illustration of D3QN algorithm for T1-UAV path planning.  \normalsize}
	\label{Fig:Algorithm_single}
\end{figure}
\subsection{Algorithm}
We have used a dueling double deep Q network (D3QN) \cite{wang2016dueling} to train the T1-UAV's optimal policy. It is worth noting that the T2-UAVs potentially use other policies to choose actions. Thus, the algorithm works in a decentralized fashion. The main D3QN algorithm is summarized and depicted in Fig. \ref{Fig:Algorithm_single}. The policy is trained for a number of episodes until convergence. In each episode, the T1-UAV navigates around other UAVs and obstacles to arrive at its destination, while collecting data from ground IoT devices. Particularly, at the beginning of each episode, the environment is reset with different settings, e.g., different starting points and destinations for the UAVs, and different locations of IoT devices. Then, at each time step, the T1-UAV observes the environment to  obtain its state. The state is parameterized and then put into a deep neural network (DNN). Using an $\epsilon$-greedy policy, the T1-UAV selects a random action with probability $\epsilon$ or follows policy greedily otherwise. When following the policy, dueling architecture is utilized. Afterwards, the UAV executes the chosen action, and then receives reward from the environment. The $\langle \mS,\mA,\mR \rangle $ tuple is saved in a replay memory, which is sampled to train the policy using back propagation. This procedure is repeated until training converges. After training, we obtain a policy, i.e., a well-trained DNN, with which the T1-UAV can perform real-time navigation.

\section{UAV Swarm Path Planning }
In this section, we consider a scenario in which a UAV swarm is assigned a mission to collect data from multiple distributed ground IoT devices, and there are also multiple T2-UAVs in the environment with unknown missions. T1-UAVs in the swarm collaborate to collect data from the IoT devices, and they communicate to share information. The goal of the UAV swarm is to find optimal paths for all T1-UAVs in order to collect the maximum amount of data while satisfying the constraints. The optimization problem for the UAV swarm can be expressed as
\begin{align}
(\text{P} 2):\argmax_{ \{\pvec^i_{t},  \forall i, \forall t \}}  & \qquad \sum_{t=0}^{T} \sum_{i=1}^{I}\sum_{n=1}^{N} q^i_{nt}\Delta t R^i_{nt} \notag  \\
\text{subject to}          &\qquad \text{time, collision avoidance, kinematic,} \notag
\\
&\qquad \text{and connectivity constraints},	\notag
\end{align}
where $i$ indicates the index of T1-UAV, $I$ is total number of T1-UAVs in the swarm, and $\pvec^i_{t}$  is the position of the $i^{th}$ UAV. The constraints in this setting include all the constraints described in Section \ref{Sec:single} plus one more constraint, which is the collision avoidance among T1-UAVs.

To solve the optimization in (P2), centralized multi-agent reinforcement learning (CMRL) or decentralized multi-agent reinforcement learning (DMRL) algorithms can be utilized. In CMRL, one computation center chooses the actions for all T1-UAVs based on the information transmitted from them. This requires communication between the center and all T1-UAVs. In real-time navigation, this communication requirement may lead to large overhead, usage of additional resources and higher delay. In addition, if one T1-UAV loses connection and is not able to send its information to the computation center, the decisions made for the other T1-UAVs may become unreliable. Besides, the policy needs to be retrained if the number of UAVs changes. Moreover, in centralized learning, the sizes of the state space and the action space will grow exponentially with the increasing number of UAVs. This will in turn increase the computational complexity significantly and reduce the learning efficiency and accuracy. As a comparison, with decentralized learning, each T1-UAV can learn its own policy and make decisions with only limited information exchange with others.  In addition, different number of UAVs can use the trained policy to make decisions without retraining. Due to these reasons,  DMRL is used to learn the optimal policies in this article.

\begin{figure}
	\centering
	\includegraphics[width=0.4\textwidth]{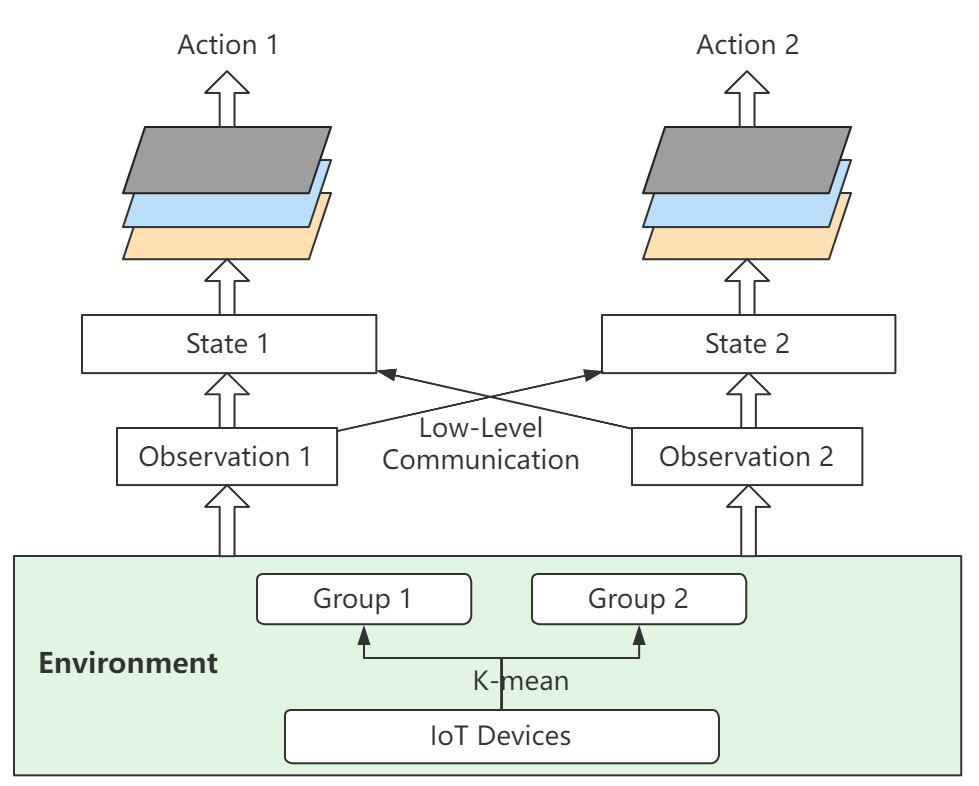}
	\caption{\small An illustration of multi-agent DMRL algorithm for path planning for two T1-UAVs performing data collection.  \normalsize}
	\label{Fig:Algorithm_multi}
\end{figure}

More specifically, D3QN with limited information exchange will be used by each T1-UAV to learn the policies. The DMRL algorithm for path planning for two T1-UAVs is depicted in Fig. \ref{Fig:Algorithm_multi}. To reduce the complexity of the problem, e.g., to manage the interactions among the T1-UAVs and reduce the risk of collision, the IoT devices are divided into groups using the K-means algorithm. One group of IoT devices is assigned to one T1-UAV. Then, each T1-UAV obtains the observations from the environment, including the observable information of T2-UAVs, information from its group of IoT devices, and its own information.  Since T1-UAVs work collaboratively, we assume that they perform low-level communication, meaning that the T1-UAVs share their full information vector with the nearby other T1-UAVs. After communication, each T1-UAV's state is constructed. Subsequently, each T1-UAV regards all other UAVs as part of the environment, and trains its policy independently according to the D3QN algorithm.  It is worth noting that, for each T1-UAV, the state and the reward function are modified based on descriptions in Section \ref{subsec:reward}. Particularly, for the $i^{th}$ T1-UAV, the information vector of the nearby T1-UAV is added to the state. Reward terms based on the minimum distance among the $i$th T1-UAV and all other T1-UAVs are added to avoid collision among them and encourage that they move relatively further away from each other.

\section{Single UAV Path Planning in the Presence of Jamming Attacks }
In this section,  we consider a scenario in which a T1-UAV needs to collect data from IoT devices. Multiple T2-UAVs also exist, one of which is an intelligent jammer that aims to attack the links between the T1-UAV and the IoT devices.  The jammer is equipped with low-cost sensors (e.g., radar) in order to sense the nearby UAVs and track the T1-UAV.

\subsection{Jammer's Path Planning Policy}
The objective of the jammer is to reduce the signal-to-interference-plus-noise ratio (SINR) of the T1-UAV subject to collision avoidance constraints, maximum travel time constraint, kinematic constraints and the start and destination constraints.  We can formulate the optimization problem as
\begin{align}
(\text{P3}):  \argmax_{ \{\pvec^J_{t},  \forall t\}}  & \qquad \E \left[\sum_{t=0}^{T^J}\sum_{n=1}^{N} q^V_{nt} \frac{1}{S^V_{nt}} \bigg| \pi^V \right] \notag  \\
\text{subject to}  &\quad \text{time, collision, and kinematic constraints,} \notag
\end{align}
where $\pvec^J_{t}$ is the position of the jammer at time $t$, $T^J$ is the total flight time of the jammer. $\pi^V$ is the decision making policy of the T1-UAV, $S^V_{nt}$ is the T1-UAV's  SINR if it is connected with the $n^{th}$ IoT device at time $t$, and  $q^V_{nt}$ is the association at time step $t$.

We can formulate the problem of the intelligent jammer trajectory design  as an MDP, and  the tuple $\langle \mS, \mA,  \mR \rangle$ is described below. Again, D3QN can be used to learn the jammer's policy.
\subsubsection{State}
In this network, the jammer can obtain the following information vectors: its own full information vector, the observable information vector of other T2-UAVs in its sensing region, the T1-UAV's observable information, the location information and mode of each IoT device;	and the available time left for the jammer. The observed information vectors can be parameterized as follows.
\begin{itemize}
	\item The first two information vectors are transformed into jammer-centric coordinates, in which the jammer's current location is the origin and the direction to the jammer's destination is the $x$-axis.
	\item The information vector of the typical UAV and the IoT nodes from the past $\tau$ time steps can be parameterized and utilized to learn the typical UAV's policy.
\end{itemize}
The parameterized information vector of the jammer can be jointly expressed as the jammer's state.
\subsubsection{Action}
Based on the jammer's kinematic constraints, permissible velocities can be sampled to build a velocity-set.  The jammer's action is the index of a velocity in the velocity-set.
\subsubsection{Reward}
The reward function of the jammer is designed based on the objective function and the constraints. Reward terms related to collision avoidance, fixed obstacle avoidance, maximum travel time constraint, and arrival-to-the-destination goal can be designed similar to those of the T1-UAV described in Section \ref{subsec:reward}. In addition, two more reward terms should be included: one is designed based on the SINR experienced at the T1-UAV (which can be learned or estimated from the UAV’s continuous reference signal received power
(RSRP) and reference signal received quality (RSRQ) reports), and the other is designed based on the distance between the jammer and the T1-UAV.

\subsection{T1-UAV's Updated Policy}
To avoid the attacks from the jammer, the T1-UAV's policy and algorithm, as well as the state and reward function design, should be updated.	Since the jammer injects interference and is generally close to the typical UAV, and the typical UAV is able to observe nearby UAVs in its sensing region, we assume that the typical UAV is able to detect the jammer. Therefore, the location information of the jammer can be obtained by the typical UAV. The jammer's locations in the past $\tau$ time steps can be used to estimate the jammer's next movement. This information can be added to the T1-UAV's state described in Section III-A. To encourage the T1-UAV to fly away from the jammer, an additional reward term is added to the original reward function in Section III-C. With the modified state and reward function, the T1-UAV's policy can be retrained using the D3QN algorithm in the presence of the intelligent jammer.

\section{Simulation Results}
In this section, we present the numerical and simulation results to demonstrate the performance of the proposed algorithms.  The considered performance metrics are the following:  1) success rate (SR), and a success means that the UAV follows a collision-free trajectory to its destination within the mission completion deadline;  2) data collection rate (DR), which is the percentage of collected data within successful missions; 3)
collision rate (CR), and 4) average path length (APL).

In simulations, the number of T2-UAVs is set to be 10. T2-UAVs use optimal reciprocal collision avoidance (ORCA) \cite{berg2011reciprocal} in choosing actions and determining their trajectories, and they do not consider collision avoidance with T1-UAVs. T1-UAVs have fixed areas for departure and landing. At the beginning of each simulation episode, the starting points and destinations of all UAVs and  the locations of IoT devices are randomly generated.

\subsection{Performance Evaluation and the Impact of the Number of T1-UAVs}
\begin{figure*}
	\centering
	\begin{minipage}{0.4\textwidth}
		\centering
		\includegraphics[width=1\textwidth]{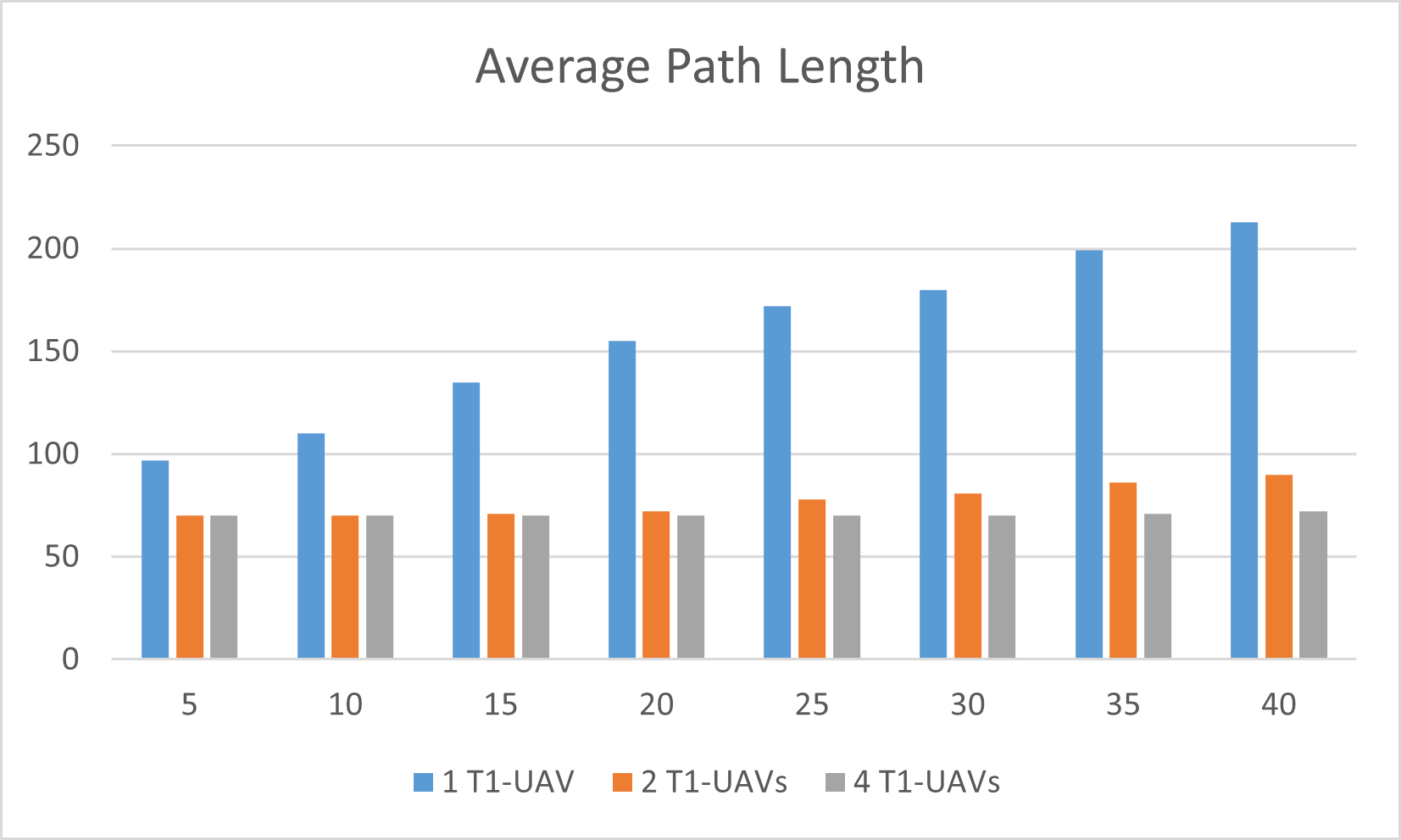} \\
		\subcaption{\scriptsize Average path length}
	\end{minipage}
	\begin{minipage}{0.4\textwidth}
		\centering
		\includegraphics[width=1\textwidth]{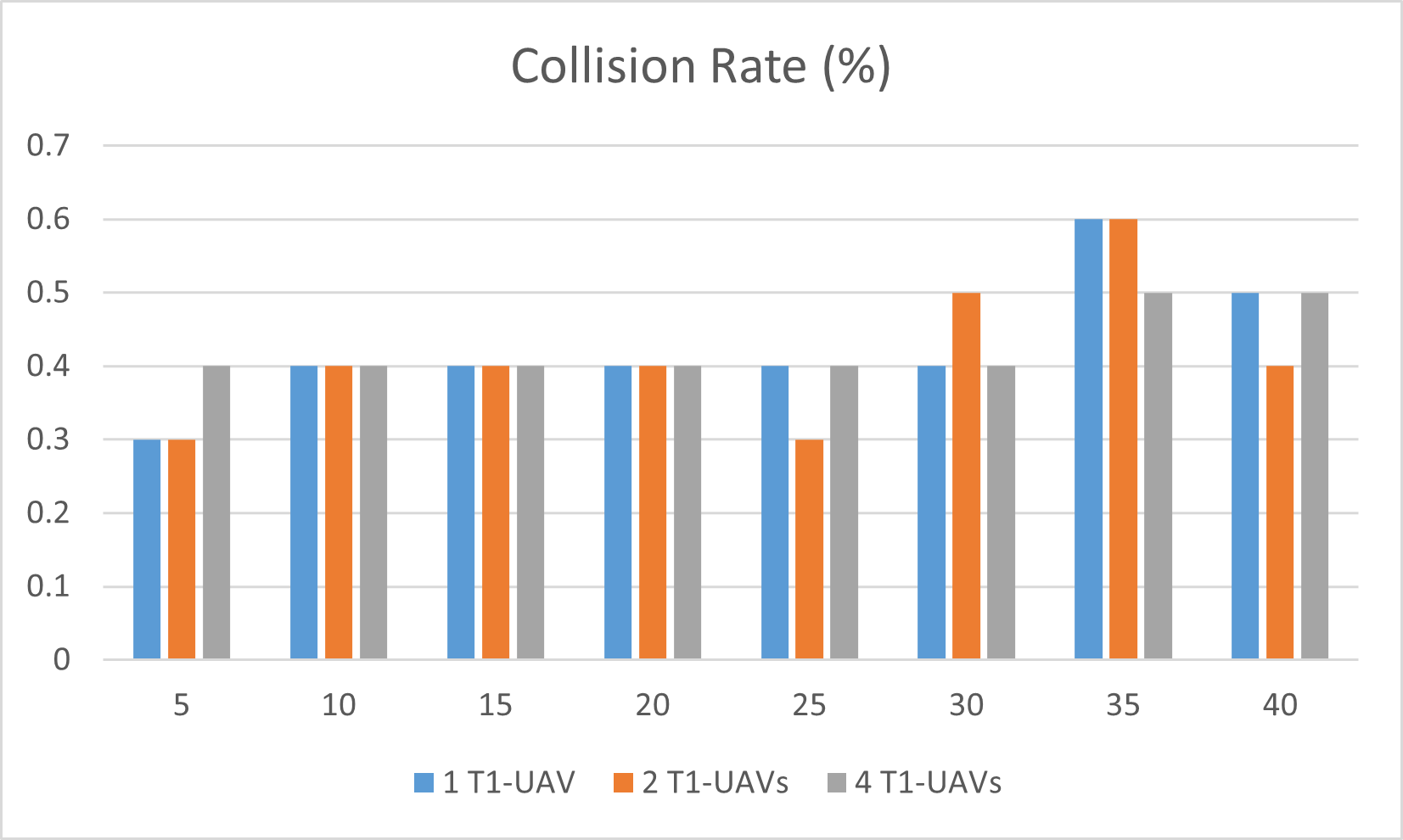} \\
		\subcaption{\scriptsize Collision rate }
	\end{minipage}
	\begin{minipage}{0.4\textwidth}
		\centering
		\includegraphics[width=1\textwidth]{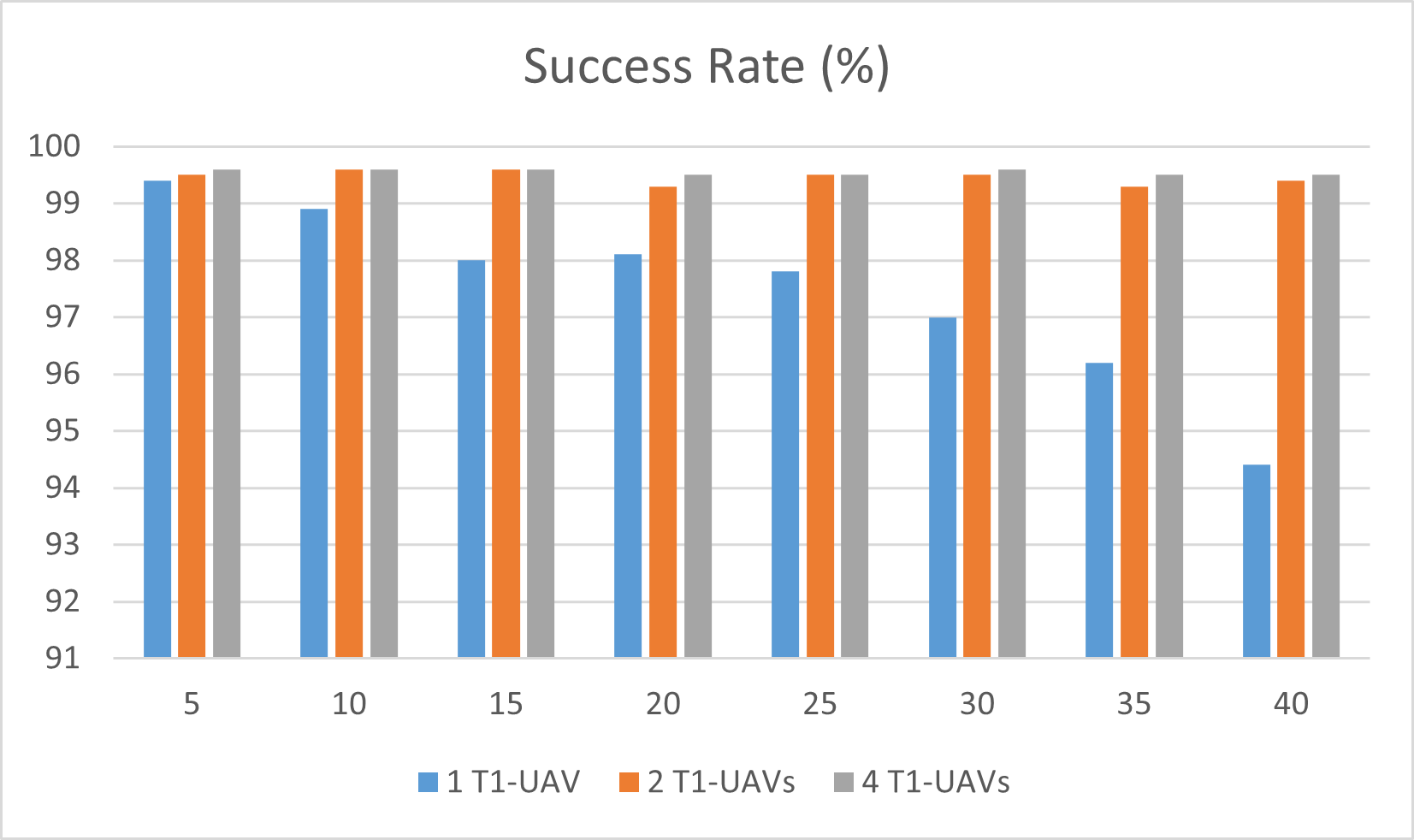} \\
		\subcaption{\scriptsize Success rate}
	\end{minipage}
	\begin{minipage}{0.4\textwidth}
		\centering
		\includegraphics[width=1\textwidth]{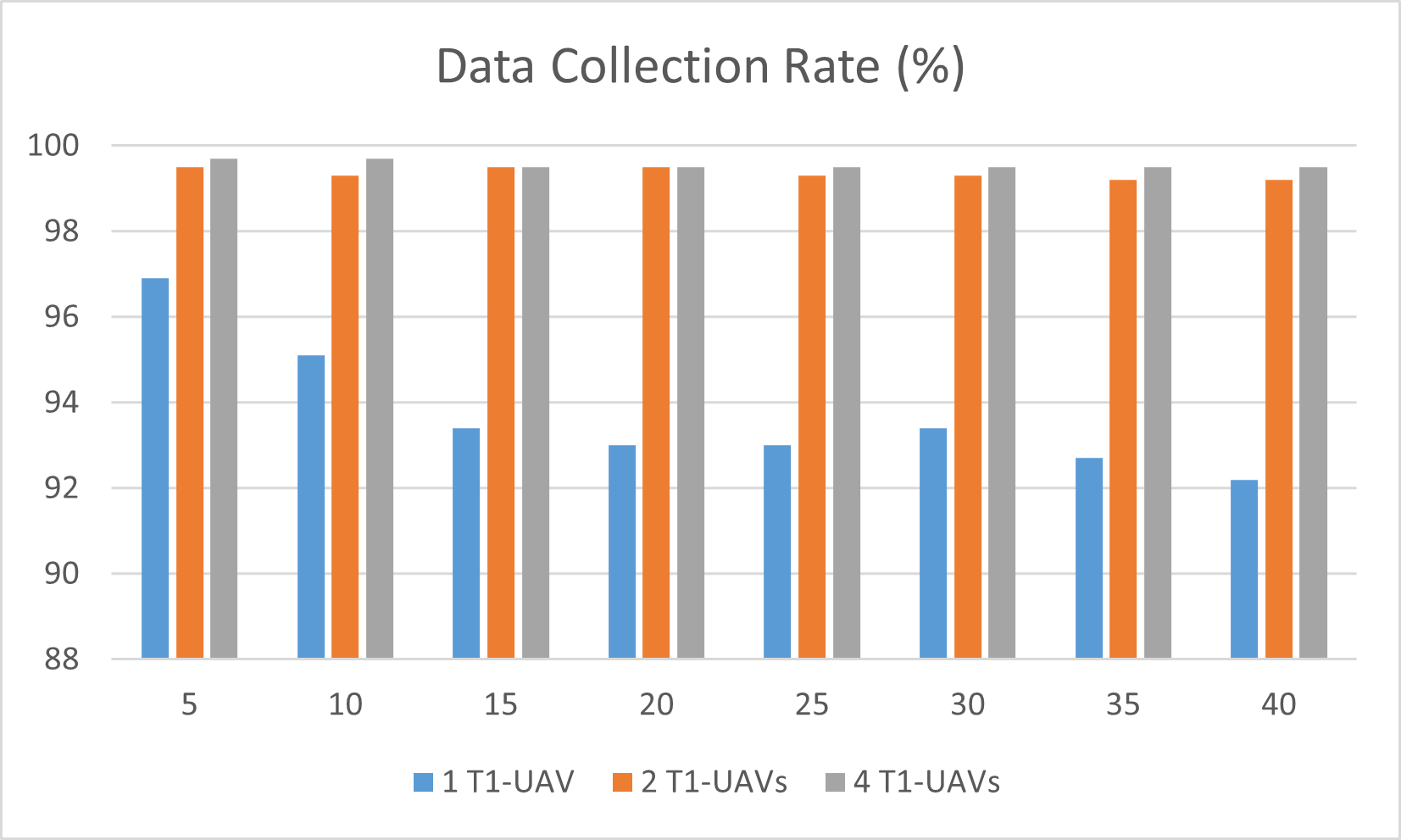}
		\subcaption{\scriptsize Data collection rate}
	\end{minipage}
	\caption{\small  APL , CR, SR, and DR for different number of IoT devices in scenarios with one, two, and four T1-UAVs.   \normalsize}
	\label{Fig:num_uav}
\end{figure*}
With the learned policy, the UAVs can perform real-time navigation. Fig. \ref{Fig:num_uav} provides the bar graphs of APL, CR, SR, and DR for different number IoT devices in scenarios involving a single T1-UAV, two T1-UAVs, and four T1-UAVs. From Fig. \ref{Fig:num_uav}(a), it can be observed that APL in the case of a single T1-UAV (denoted by APL1) is the largest, APL in scenario with two T1-UAVs (denoted by APL2) is smaller, and the smallest APL is observed in the scenario with four T1-UAVs (denoted by APL4). The reason is that if there are more T1-UAVs working collaboratively, each UAV only needs to collect data from a smaller number of IoT nodes, and correspondingly requires less time and shorter path to complete the mission. We further notice that  APL2 is not much larger than APL4, due to the reason that even though each UAV needs to collect data from a smaller number of devices, the UAVs still need to fly from their starting points to their destinations, resulting in a certain minimum flight duration. Overall, Fig. \ref{Fig:num_uav}(a) demonstrates that increasing the number of T1-UAVs can reduce the UAV path length, but the APL levels off at a certain minimum value.
In addition, Fig. \ref{Fig:num_uav}(a) shows that APL1 grows when the number of IoT devices, $N$, is increased. The reason is that the UAV needs to fly relatively close to each IoT node to be able to collect data, and thus if more IoT nodes need to upload data, the T1-UAV needs to stay in the area for a longer time duration in order to complete its mission.  APL2 also increases with increasing $N$, but the increment is much smaller than that in the scenario with a single T1-UAV. And APL4 rises very slightly with the increasing $N$. Fig. \ref{Fig:num_uav}(b) shows that overall the collision rate is smaller than 0.6\%, indicating that the proposed algorithms help the UAVs successfully avoid collisions. We note that the CR rises slightly with increasing $N$, since flying over a longer duration or path to complete the mission leads to higher risk of collisions.

In Fig. \ref{Fig:num_uav}(c), we notice that in the scenario with a single T1-UAV, SR is above 94\%  even when there are 40 IoT devices in the environment. The SR grows further when the number of IoT devices is decreased. This is due to the fact that if $N$ decreases, APL1 is reduced, and thus it becomes easier for the UAV to satisfy its mission completion deadline, and correspondingly it achieves a higher success rate. In addition, Fig. \ref{Fig:num_uav}(c) shows that in scenarios with two and four T1-UAVs, the SRs are above 99\%, indicating that almost all UAVs can execute successful trajectories.  Fig. \ref{Fig:num_uav}(d) shows that DR in the scenario with a single T1-UAV  decreases with increasing $N$, since it becomes easier to miss some data if there are more devices to collect data from. However, DRs in scenarios with two and four T1-UAVs are above 99\%. From Figs. \ref{Fig:num_uav}(c) and \ref{Fig:num_uav}(d), we can conclude that having two or four T1-UAVs lead to significantly enhanced performance levels (compared to the scenario with a single T1-UAV) with over 99\% success rate and over 99\% data collection rate.

\subsection{Trajectories and Performance in the Presence of Jamming Attacks}
\begin{figure*}
	\centering
	\begin{minipage}{0.3\textwidth}
		\centering
		\includegraphics[width=1\textwidth]{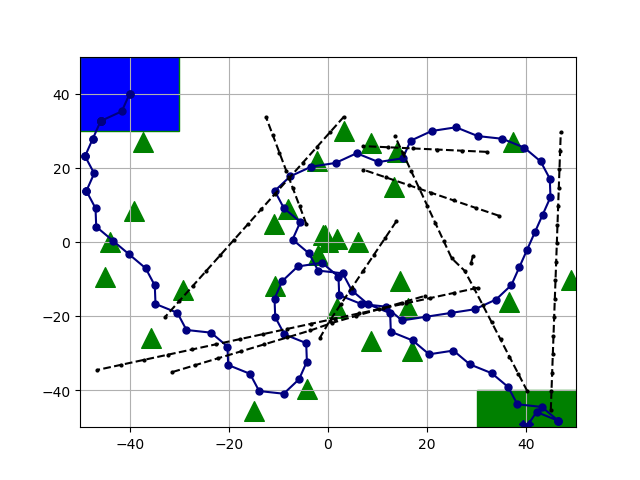}
		\subcaption{\scriptsize Single T1-UAV path planning}
	\end{minipage}
	\begin{minipage}{0.3\textwidth}
		\centering
		\includegraphics[width=1\textwidth]{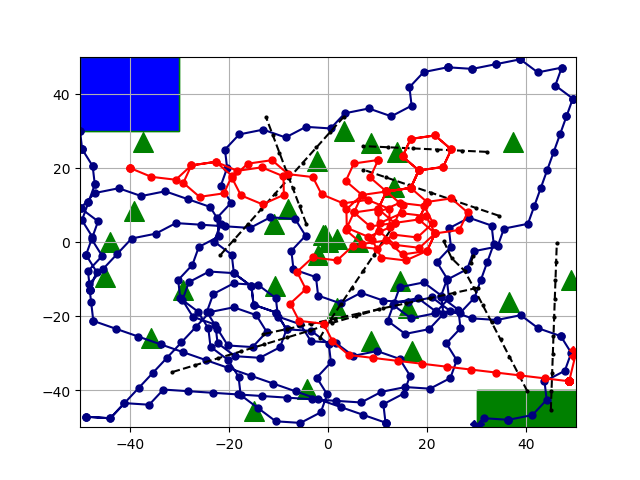}
		\subcaption{\scriptsize Single 	T1-UAV path planning in the presence of a jammer}
	\end{minipage}
	\begin{minipage}{0.3\textwidth}
		\centering
		\includegraphics[width=1\textwidth]{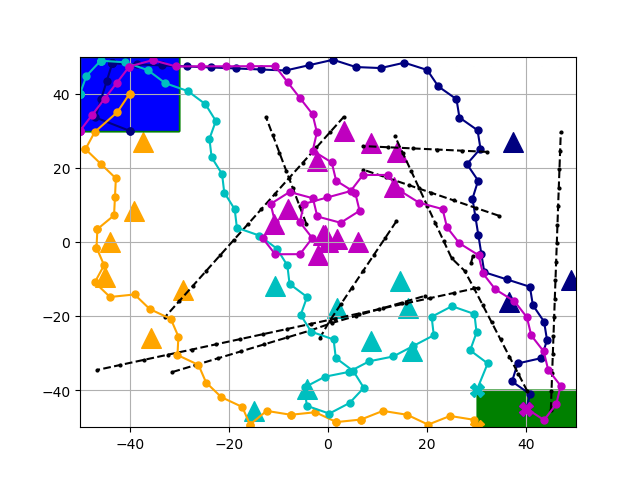}
		\subcaption{\scriptsize Path planning for four T1-UAVs}
	\end{minipage}
	\caption{\small Examples of UAV trajectories in different scenarios. }
	\label{Fig:intelligent_jammer_traj}
\end{figure*}

In Fig. \ref{Fig:intelligent_jammer_traj}, we plot the single T1-UAV trajectory, single T1-UAV trajectory in the presence of the intelligent mobile jammer, and trajectories when there are four T1-UAVs. In the illustrations, the departure and landing areas for the T1-UAVs are displayed by blue and green areas, respectively. The trajectories of T2-UAVs in the environment are depicted by black dashed lines with dots.
In Figs. \ref{Fig:intelligent_jammer_traj}(a) and \ref{Fig:intelligent_jammer_traj}(b), the trajectories of the T1-UAV are presented by navy lines with dots, and the trajectory of the jammer is shown in red lines with dots.
The IoT nodes are marked by green triangles. When comparing the trajectories in Figs. \ref{Fig:intelligent_jammer_traj}(a) and \ref{Fig:intelligent_jammer_traj}(b), we observe that the existence of the jammer has significant influence on the T1-UAV, specifically leading to a very curvy and long T1-UAV trajectory. In such a case, it becomes easier for the T1-UAV to violate its mission completion deadline and fail to collect data from some IoT devices. Correspondingly, SR and DR will be decreased. On the other hand, we can observe that the T1-UAV trajectory is curvy since it tries to go further away from the jammer due to the impact of the reward terms related to the distance between the T1-UAV and the jammer. With this, the T1-UAV can reduce the collision risk with the jammer and lower the interference from the jammer. In addition, the T1-UAV can return to the IoT devices if the jammer leaves or ceases to operate for a certain duration of time (e.g., due to energy/power budget constraints) with the goal to collect data and recover the SR and DR to a certain degree. 

In Fig. \ref{Fig:intelligent_jammer_traj}(c), the IoT devices are divided into four groups and marked by different colors. Each T1-UAV is assigned to collect data from one group, and its trajectory is denoted by lines in the same color with the IoT devices. The figure shows that each T1-UAV flies towards its own group of IoT devices, collects data, and finally arrives at its destination. Comparing Figs. \ref{Fig:intelligent_jammer_traj}(a) and \ref{Fig:intelligent_jammer_traj}(c), we  observe that the trajectories in Fig. \ref{Fig:intelligent_jammer_traj}(c) are much shorter than that in \ref{Fig:intelligent_jammer_traj}(a), indicating that more T1-UAVs working collaboratively can reduce the path length and correspondingly the mission completion time.

\section{Conclusion}

In this article, we have investigated robust and decentralized UAV path planning in IoT networks in the presence of other noncooperative UAVs and a jammer UAV. We have considered three different realistic scenarios, including single UAV path planning, UAV swarm path planning, and single UAV path planning in the presence of an intelligent jammer. We have formulated the optimization problem in each case, and converted the problem into an MDP with well-designed state, action, and reward functions. We have proposed decentralized multi-agent reinforcement learning algorithms for UAV path planning in these three scenarios under practical constraints. The simulation results demonstrate that with learning-based path planning, the UAVs can complete their missions with high success rates and data collection rates. In addition, the UAVs are able to execute different trajectories as a defensive measure against the intelligent jammer.

\bibliographystyle{IEEEtran}
\bibliography{6G_ml}

\begin{thebibliography}{10}
\providecommand{\url}[1]{#1}
\csname url@samestyle\endcsname
\providecommand{\newblock}{\relax}
\providecommand{\bibinfo}[2]{#2}
\providecommand{\BIBentrySTDinterwordspacing}{\spaceskip=0pt\relax}
\providecommand{\BIBentryALTinterwordstretchfactor}{4}
\providecommand{\BIBentryALTinterwordspacing}{\spaceskip=\fontdimen2\font plus
\BIBentryALTinterwordstretchfactor\fontdimen3\font minus
  \fontdimen4\font\relax}
\providecommand{\BIBforeignlanguage}[2]{{%
\expandafter\ifx\csname l@#1\endcsname\relax
\typeout{** WARNING: IEEEtran.bst: No hyphenation pattern has been}%
\typeout{** loaded for the language `#1'. Using the pattern for}%
\typeout{** the default language instead.}%
\else
\language=\csname l@#1\endcsname
\fi
#2}}
\providecommand{\BIBdecl}{\relax}
\BIBdecl

\bibitem{yang20196g}
P.~Yang, Y.~Xiao, M.~Xiao, and S.~Li, ``{6G} wireless communications: Vision
  and potential techniques,'' \emph{IEEE network}, vol.~33, no.~4, pp. 70--75,
  2019.

\bibitem{mao2021optimizing}
B.~Mao, F.~Tang, Y.~Kawamoto, and N.~Kato, ``Optimizing computation offloading
  in satellite-{UAV}-served {6G IoT}: a deep learning approach,'' \emph{IEEE
  Network}, vol.~35, no.~4, pp. 102--108, 2021.

\bibitem{li2018learning}
H.~Li, K.~Ota, and M.~Dong, ``Learning {IoT} in edge: Deep learning for the
  {I}nternet of {T}hings with edge computing,'' \emph{IEEE network}, vol.~32,
  no.~1, pp. 96--101, 2018.

\bibitem{you2021towards}
X.~You, C.-X. Wang, J.~Huang, X.~Gao, Z.~Zhang, M.~Wang, Y.~Huang, C.~Zhang,
  Y.~Jiang, J.~Wang \emph{et~al.}, ``Towards {6G} wireless communication
  networks: Vision, enabling technologies, and new paradigm shifts,''
  \emph{Science China Information Sciences}, vol.~64, no.~1, pp. 1--74, 2021.

\bibitem{araniti2021toward}
G.~Araniti, A.~Iera, S.~Pizzi, and F.~Rinaldi, ``Toward {6G} non-terrestrial
  networks,'' \emph{IEEE Network}, 2021.

\bibitem{saad2019vision}
W.~Saad, M.~Bennis, and M.~Chen, ``A vision of {6G} wireless systems:
  Applications, trends, technologies, and open research problems,'' \emph{IEEE
  network}, vol.~34, no.~3, pp. 134--142, 2019.

\bibitem{yang2020artificial}
H.~Yang, A.~Alphones, Z.~Xiong, D.~Niyato, J.~Zhao, and K.~Wu,
  ``Artificial-intelligence-enabled intelligent {6G} networks,'' \emph{IEEE
  Network}, vol.~34, no.~6, pp. 272--280, 2020.

\bibitem{liu2019trajectory}
X.~Liu, Y.~Liu, Y.~Chen, and L.~Hanzo, ``Trajectory design and power control
  for multi-{UAV} assisted wireless networks: A machine learning approach,''
  \emph{IEEE Transactions on Vehicular Technology}, vol.~68, no.~8, pp.
  7957--7969, 2019.

\bibitem{bayerlein2021multi}
H.~Bayerlein, M.~Theile, M.~Caccamo, and D.~Gesbert, ``Multi-{UAV} path
  planning for wireless data harvesting with deep reinforcement learning,''
  \emph{IEEE Open Journal of the Communications Society}, vol.~2, pp.
  1171--1187, 2021.

\bibitem{li2021drlr}
T.~Li, W.~Liu, Z.~Zeng, and N.~Xiong, ``{DRLR}: A deep reinforcement learning
  based recruitment scheme for massive data collections in {6G}-based {IoT}
  networks,'' \emph{IEEE Internet of Things journal}, 2021.

\bibitem{chen2020autonomous}
Y.-J. Chen, D.-K. Chang, and C.~Zhang, ``Autonomous tracking using a swarm of
  {UAV}s: A constrained multi-agent reinforcement learning approach,''
  \emph{IEEE Transactions on Vehicular Technology}, vol.~69, no.~11, pp.
  13\,702--13\,717, 2020.

\bibitem{hu2021distributed}
Y.~Hu, M.~Chen, W.~Saad, H.~V. Poor, and S.~Cui, ``Distributed multi-agent meta
  learning for trajectory design in wireless drone networks,'' \emph{IEEE
  Journal on Selected Areas in Communications}, vol.~39, no.~10, pp.
  3177--3192, 2021.

\bibitem{zhong2021multi}
R.~Zhong, X.~Liu, Y.~Liu, and Y.~Chen, ``Multi-agent reinforcement learning in
  {NOMA}-aided uav networks for cellular offloading,'' \emph{IEEE Transactions
  on Wireless Communications}, 2021.

\bibitem{wang2016dueling}
Z.~Wang, T.~Schaul, M.~Hessel, H.~Hasselt, M.~Lanctot, and N.~Freitas,
  ``Dueling network architectures for deep reinforcement learning,'' in
  \emph{International conference on machine learning}.\hskip 1em plus 0.5em
  minus 0.4em\relax PMLR, 2016, pp. 1995--2003.

\bibitem{berg2011reciprocal}
J.~v.~d. Berg, S.~J. Guy, M.~Lin, and D.~Manocha, ``Reciprocal n-body collision
  avoidance,'' in \emph{Robotics research}.\hskip 1em plus 0.5em minus
  0.4em\relax Springer, 2011, pp. 3--19.

\end{thebibliography}

\end{document}